\documentclass[twocolumn,aps,superscriptaddress,showpacs,nofootinbib]{revtex4}

\usepackage{amssymb}
\usepackage{amsmath}
\usepackage{graphicx}
\usepackage[normalem]{ulem}
\usepackage[dvips]{color}

\begin{document}
\title{Elliptic flow splitting as a probe of the QCD phase
structure at finite baryon chemical potential}

\author{Jun Xu}
\email{xujun@sinap.ac.cn} \affiliation{Shanghai Institute of Applied
Physics, Chinese Academy of Sciences, Shanghai 201800, China}

\author{Taesoo Song}
\affiliation{Cyclotron Institute and Department of Physics and
Astronomy, Texas A$\&$M University, College Station, Texas 77843,
USA}

\author{Che Ming Ko}
\affiliation{Cyclotron Institute and
Department of Physics and Astronomy, Texas A$\&$M University,
College Station, Texas 77843, USA}

\author{Feng Li}
\affiliation{Cyclotron Institute and
Department of Physics and Astronomy, Texas A$\&$M University,
College Station, Texas 77843, USA}

\date{\today}

\begin{abstract}
Using a partonic transport model based on the 3-flavor
Nambu-Jona-Lasinio model and a relativistic hadronic transport model
to describe, respectively, the evolution of the initial partonic and
the final hadronic phase of heavy-ion collisions at energies carried
out in the Beam-Energy Scan program of the Relativistic Heavy Ion
Collider, we have studied the effects of both the partonic and
hadronic mean-field potentials on the elliptic flow of particles
relative to that of their antiparticles. We find that to reproduce
the measured relative elliptic flow differences between nucleons and
antinucleons as well as between kaons and antikaons requires a
vector coupling constant as large as 0.5 to 1.1 times the scalar
coupling constant in the Nambu-Jona-Lasinio model. Implications of
our results in understanding the QCD phase structure at finite
baryon chemical potential are discussed.
\end{abstract}


\maketitle

The main purpose of the experiments involving collisions of heavy
nuclei at relativistic energies is to study the properties of
produced quark-gluon plasma (QGP) and its phase transition to
hadrons. It is known from the lattice Quantum Chromodynamics (QCD)
that for QGP of small baryon chemical potential, such as that formed
at top energies of the Relativistic Heavy Ion Collider (RHIC) and
the Large Hadron Collider (LHC), the hadron-quark phase transition
(HQPT) is a smooth crossover~\cite{Ber05,Baz12}. For QGP of finite
baryon chemical potential produced at lower collision energies,
studies based on various theoretical models have indicated, however,
that the HQPT is expected to change to a first-order
one~\cite{Asa89,Fuk08,Car10,Wei12}. To determine if the critical
point, at which the crossover HQPT changes to a first-order one,
exists and where it is located in the QCD phase diagram is important
for understanding the phase structure of QCD and thus the nature of
the strong interaction. To search for the critical point, the
Beam-Energy Scan (BES) program has been carried out at RHIC to look
for its signals at lower collision energies of
$\sqrt{s_{NN}}=7.7\sim39$ GeV.

Although there is no definitive conclusion on the existence or the
location of the critical point, many interesting phenomena different
from those at higher collision energies have been
observed~\cite{Kum11,Moh11}. Among them is the increasing splitting
between the elliptic flow ($v_2$) of particles, i.e, the second
Fourier coefficient in the azimuthal distribution of their momenta
in the plane perpendicular to the participant or reaction plane, and
that of their antiparticles with decreasing collision
energy~\cite{STAR13}. This result also indicates the breakdown of
the number of constituent quark scaling of $v_2$~\cite{STAR07} at
lower collision energies. The latter states that the scaled elliptic
flow, which is obtained from dividing the hadron elliptic flow by
its number of constituent quarks as a function of a similarly scaled
transverse kinetic energy, is similar for all hadrons, and this has
been considered as an evidence for the existence of QGP in heavy-ion
collisions at higher energies. Various explanations have been
proposed to account for the observed splitting of particle and
antiparticle $v_2$. In Ref.~\cite{Bur11}, it was suggested that in
the presence of the strong magnetic field in non-central heavy-ion
collisions, the chiral magnetic wave induced by the axial anomaly in
QCD can generate an electric quadrupole moment in the produced
baryon-rich QGP. This can then lead to the splitting of the $v_2$ of
oppositely charged particles and antiparticles, particularly that of
$\pi^+$ and $\pi^-$ due to their similar final-state interactions in
the hadronic matter. The $v_2$ splitting of particles and
antiparticles may also be attributed to different $v_2$ of
transported and produced partons~\cite{Dun11}, different rapidity
distributions of quarks and antiquarks~\cite{Gre12}, and the
conservation of baryon charge, strangeness, and
isospin~\cite{Ste12}.

On the other hand, we have shown in our previous studies that the
different mean-field potentials for hadrons and
antihadrons~\cite{Xu12} or quarks and antiquarks~\cite{Son12} in the
baryon-rich matter produced at lower collision
energies can describe qualitatively the $v_2$ splitting of particles
and their antiparticles. This is due to the fact that particles with
attractive potentials are more likely to be trapped in the system
and move in the direction perpendicular to the participant plane,
while those with repulsive potentials are more likely to leave the
system and move along the participant plane, thus reducing and
enhancing their respective elliptic flows. However, the relative
$v_2$ difference between $p$ and $\bar{p}$ was underestimated in
both our previous studies, and that between $K^+$ and $K^-$ was
overestimated in Ref.~\cite{Xu12} with only hadronic potentials and
underestimated in Ref.~\cite{Son12} with only partonic potentials.

In the present study, we include both the partonic potentials from
the Nambu-Jona-Lasinio (NJL) model~\cite{NJL61} and the hadronic
potentials that were known in the literature on subthreshold
particle production in heavy-ion collisions~\cite{GQL94,GQL97}. To
reproduce quantitatively the relative $v_2$ difference for $p$ and
$\bar{p}$ as well as $K^+$ and $K^-$ in mini-bias Au+Au collisions
at $\sqrt{s_{NN}}=7.7$ GeV, we find that the ratio $R_V$ of the
vector coupling $G_V$ to the scalar-pseudoscalar coupling $G$ in the
NJL model is constrained to values between $0.5$ and $1.1$. It has
been shown that this ratio determines the density beyond which the
quark matter is formed in hybrid stars~\cite{Sha13}, which is
important in understanding the composition of the recently
discovered two-solar-mass neutron star~\cite{Dem10}. More
importantly, studies based on both the NJL model and the
Polyakov-Nambu-Jona-Lasinio (PNJL)
model~\cite{Asa89,Fuk08,Car10,Wei12} have shown that the existence
and the location of the HQPT critical point in the QCD phase diagram
depends on the value of $R_V$. Our results thus suggest that
studying the $v_2$ splitting of particles and their antiparticles
provides the possibility of mapping out the QCD phase diagram at
finite baryon chemical potential and to better understand the nature
of the strong interaction.

We include the mean-field potential effect by extending the
string-melting version of a multiphase transport (AMPT)
model~\cite{Lin05}, which has been successfully used to describe the
harmonic flows and the di-hadron correlation at both RHIC and
LHC~\cite{Xu11}. In this model, the initial hadrons, which are
generated by the Heavy-Ion Jet Interaction Generator (HIJING)
model~\cite{Xnw91} via the Lund string fragmentation model, are
converted to their constituent or valence quarks and antiquarks.
Different from previous studies at higher energies, the evolution of
partons in time and space is modeled by a 3-flavor NJL transport
model~\cite{Son12}. The NJL Lagrangian is written as~\cite{Wei12}
\begin{eqnarray}
\mathcal{L}&=&\bar{\psi}(i\not{\partial}-M)\psi+\frac{G}{2}\sum_{a=0}^{8}\bigg[(\bar{\psi}\lambda^a\psi)^2+(\bar{\psi}i\gamma_5\lambda^a\psi)^2\bigg]\nonumber\\
&+&\sum_{a=0}^{8}\bigg[\frac{G_V}{2}(\bar{\psi}\gamma_\mu\lambda^a\psi)^2+\frac{G_A}{2}(\bar{\psi}\gamma_\mu\gamma_5\lambda^a\psi)^2\bigg]\nonumber\\
&-&K\bigg[{\rm det}_f\bigg(\bar{\psi}(1+\gamma_5)\psi\bigg)+{\rm
det}_f\bigg(\bar{\psi}(1-\gamma_5)\psi\bigg)\bigg],
\end{eqnarray}
with the quark field $\psi=(\psi_u, \psi_d, \psi_s)^T$, the current
quark mass matrix $M={\rm diag}(m_u, m_d, m_s)$, and the Gell-Mann
matrices $\lambda^{a}$ in $SU(3)$ flavor space. In the case that the
vector and axial-vector interactions are generated by the Fierz
transformation of the scalar and pseudo-scalar interactions, their
coupling strengths are given by $G_V=G_A=G/2$, while $G_V=1.1G$ was
used in Ref.~\cite{Wei92} to give a better description of the vector
meson-mass spectrum based on the NJL model. Other parameters are
taken from Refs.~\cite{Wei92,Wei12} as $m_u=m_d=3.6$ MeV, $m_s=87$
MeV, $G\Lambda^2=3.6$, $K\Lambda^{5}=8.9$, and
$\Lambda=750$ MeV is the cut-off value in the momentum integration.
In the mean-field approximation, the above Lagrangian leads to an
attractive scalar mean-field potential for both quarks and
antiquarks~\cite{Son12}. With a nonvanishing $G_V$, it further gives
rise to a repulsive vector mean-field potential for quarks but an
attractive one for antiquarks in a baryon-rich quark matter.

For the scattering cross sections between quarks and antiquarks, we
assume that they are isotropic and have a constant value that is
determined from fitting the measured charged-particle elliptic flow.
The quark matter then evolves under the influence of both mean-field
potentials and two-body scatterings until the chiral symmetry is
broken, i.e., the effective mass of light quarks, which is
determined by Eq.~(4) in Ref.~\cite{Son12}, is larger than about
$200$ MeV.

At hadronization, quarks and antiquarks in the AMPT model are
converted to hadrons via a spatial coalescence model by considering
the invariant mass of nearest quarks and antiquarks and converting
them into a hadron with the closest mass. This is different from the
one used in Ref.~\cite{Son12} based on the phase space distribution
of quarks and antiquarks and the Wigner functions of produced
hadrons.  Although the spatial coalescence is more schematic, it has
been shown to give a reasonable description of many experimental
data.

For the scatterings between hadrons in the hadronic stage, they are
described by a relativistic transport (ART) model~\cite{Bal95} that
has been extended to also include particle-antiparticle
annihilations and their inverse reactions. For the hadronic
potentials, they are included as in our previous work~\cite{Xu12}
using the phenomenologically determined relativistic mean-field
model~\cite{GQL94} for nucleons and antinucleons, and effective
chiral Lagrangian~\cite{GQL97} for kaons and antikaons. Due to the
G-parity invariance, the potential for antinucleons is much more
attractive than that for nucleons, and that for kaons is slightly
repulsive while that for antikaons is deeply attractive in a
baryon-rich hadronic matter.

For the centrality of heavy-ion collisions, we use the empirical
formula $c=\pi \text{b}^2/\sigma_{in}$~\cite{Bro02} to determine the
relation between the centrality $c$ used in the experimental
analysis and the impact parameter b, where the total nucleus-nucleus
inelastic cross section is $\sigma_{in}\approx 686$ fm$^2$ from the
Glauber model calculation using the nucleon-nucleon inelastic cross
section of about $30.8$ mb at $\sqrt{s_{NN}}=7.7$ GeV from
Ref.~\cite{STAR12}. The mean-field potentials in both the partonic
phase and the hadronic phase are then calculated using the test
particle method~\cite{Won82} with parallel events that correspond to
the same impact parameter. We note that although the initial parton
distribution obtained from AMPT includes fluctuations, they are
largely destroyed when using the test particle method to calculate
the mean fields. For the elliptic flow addressed in the present
study, this is, however, not an important effect as it is mostly due
to the collision geometry and the fluctuation of initial
eccentricity is not large.

\begin{figure}[h]
\centerline{\includegraphics[scale=0.8]{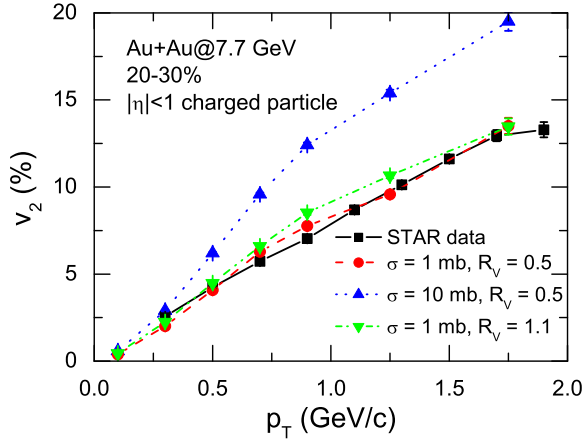}} \caption{(Color
online) Transverse momentum dependence of the elliptic flow of
mid-pseudorapidity charged particles in mid-central Au+Au collisions
at $\sqrt{s_{NN}}=7.7$ GeV for different values of the parton
scattering cross section $\sigma$ and the ratio $R_V$ of the vector
coupling constant $G_V$ to the scalar coupling constant $G$ in the
NJL model. The experimental data from the STAR Collaboration is from
Ref.~\cite{STAR12}.} \label{fit}
\end{figure}

To fix the parton scattering cross section in the NJL transport
model, we compare the transverse momentum ($p_T$) dependence of the
elliptic flow of mid-pseudorapidity charged particles measured in
mid-central Au+Au collisions at $\sqrt{s_{NN}}=7.7$ GeV with results
obtained for different values of the parton scattering cross section
$\sigma$, which is further taken to be isotropic. The elliptic flow
is calculated from the average azimuthal anisotropy of particles
with respect to the participant plane, i.e., $v_2=\langle
\cos[2(\phi-\Psi_2)] \rangle $, where $\phi=\rm atan2(p_y,p_x)$ is
the azimuthal angle of the particle momentum at the final stage, and
$\Psi_2=[\rm atan2(\langle r_p^2\sin2\phi_p\rangle,\langle
r_p^2\cos2\phi_p\rangle)+\pi]/2$ is the azimuthal angle of the
participant plane at the initial stage with $r_p$ and $\phi_p$ being
the polar coordinates of the participants.  As shown in
Fig.~\ref{fit}, the $v_2$ of charged hadrons is larger for a larger
parton scattering cross section, and we find that the experimentally
measured $v_2$ of charged particles from the STAR
Collaboration~\cite{STAR12} can be reasonably reproduced with a
parton scattering cross section of $\sigma=1$ mb. This value is
about half the average of quark-quark and quark-antiquark elastic
cross sections calculated from the NJL model~\cite{Marty:2013ita}.
The magnitude of $v_2$ is, however, rather insensitive to the value
of $R_V$, as increasing $R_V$ from 0.5 to 1.1 only increases
slightly the value of $v_2$. These results remain similar if
different methods are used to calculate the hadron $v_2$. We note
that the small parton cross section required for describing the
observed charged particle elliptic flow partially accounts for the
fact that treating particles in the corona of a heavy ion collision
as a partonic matter overestimates the elliptic flow.

\begin{figure}[h]
\centerline{\includegraphics[scale=0.8]{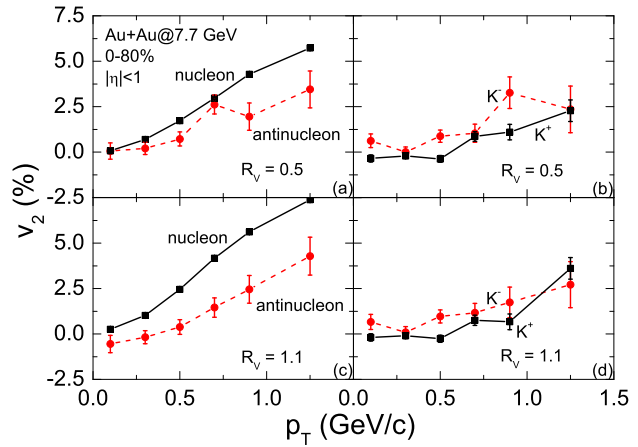}}
\caption{(Color online) Transverse momentum dependence of the
initial elliptic flows of mid-pseudorapidity nucleons and kaons as
well as their antiparticles right after hadronization in mini-bias
Au+Au collisions at $\sqrt{s_{NN}}=7.7$ GeV for different values of
$R_V=G_V/G$ in the NJL model.} \label{v2pt_ini}
\end{figure}

Figure~\ref{v2pt_ini} displays the $p_T$ dependence of initial $v_2$
for mid-pseudorapidity nucleons and kaons as well as their
antiparticles right after hadronization in mini-bias Au+Au
collisions at $\sqrt{s_{NN}}=7.7$ GeV, with the parton scattering
cross section of 1 mb as determined above and for two values of
$R_V=0.5$ and $1.1$. It is seen that in both cases, the elliptic
flow is larger for nucleons than for antinucleons and for $K^-$ than
for $K^+$, and the difference is larger for the larger value of
$R_V$, especially for nucleons and anitnucleons. These results are
qualitatively consistent with those in Ref.~\cite{Son12}, although
different parton scattering cross sections and hadronization
criteria are used. The larger nucleon than antinucleon elliptic flow
can be understood from the opposite effects of the partonic vector
potential on quarks and antiquarks, which lead to a larger quark
than antiquark elliptic flow. The reason for the larger $K^-(\bar
us)$ than $K^+(u\bar s)$ elliptic flow is, however, more
complicated, as both consist of a quark and an antiquark. As shown
in Fig.~3 of Ref.~\cite{Son12}, because of the vector potential,
which affects light quarks more than strange quarks as a result of
the small light quark mass, the light quark elliptic flow increases
faster with time than that of the strange quark. However, while the
elliptic flow of strange quarks continues to increase with time,
that of light quarks decreases at later times due to the stronger
attractive scalar potential for light quarks than for strange
quarks. As a result, the $K^-$ elliptic flow can be larger than the
$K^+$ elliptic flow after their production from quark coalescence.

\begin{figure}[h]
\centerline{\includegraphics[scale=0.8]{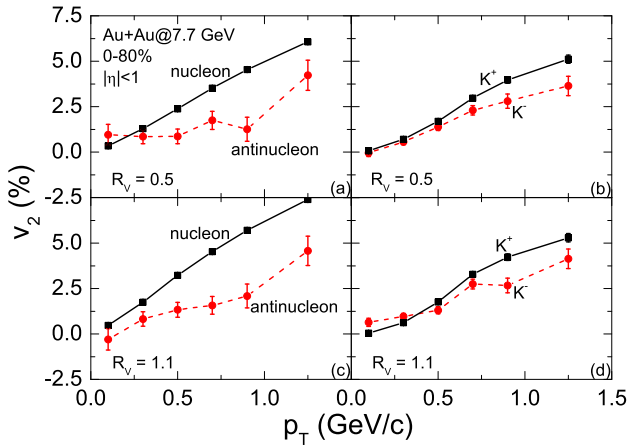}}
\caption{(Color online) Same as Fig.~\ref{v2pt_ini} but for results
after hadronic evolution.} \label{v2pt_final}
\end{figure}

The final differential elliptic flows after the evolution of the
hadronic phase are shown in Fig.~\ref{v2pt_final}, and they are
larger than their corresponding values in the beginning of the
hadronic stage. Because of the repulsive potential for nucleons and
the attractive potential for antinucleons in the baryon-rich
hadronic matter, the $v_2$ of nucleons, which is larger initially,
remains larger than that of antinucleons after hadronic evolution.
For the $v_2$ of $K^+$ and $K^-$, which is initially larger for
$K^-$ than for $K^+$, the ordering is, on the other hand, reversed
by the repulsive $K^+$ and attractive $K^-$ potential in the
baryon-rich hadronic matter, resulting in a larger $v_2$ for $K^+$
than for $K^-$ after hadronic evolution. These effects are seen for
both values of $R_V$.

\begin{figure}[h]
\centerline{\includegraphics[scale=0.8]{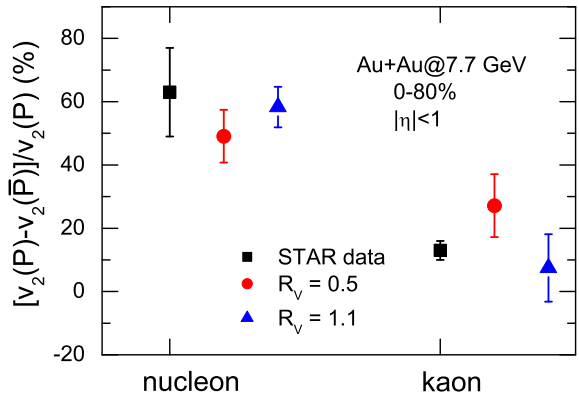}}
\caption{(Color online) Relative elliptic flow difference between
nucleons and antinucleons as well as kaons and antikaons for
different values of $R_V=G_V/G$ in the NJL model compared with the
STAR data~\cite{Moh11}.} \label{ratio_final}
\end{figure}

Figure~\ref{ratio_final} compares the $p_T$-integrated relative
elliptic flow differences $[v_2({\rm P})-v_2({\rm\bar P})]/v_2({\rm
P})$ between nucleons and antinucleons as well as between kaons and
antikaons for different values of $R_V$ with the experimental
results from the STAR Collaboration~\cite{Moh11}. The STAR results
can now be quantitatively reproduced with both $R_V=0.5$ and
$R_V=1.1$ within the statistical error. Also, we find that with
increasing value of $R_V$ the relative $v_2$ difference between
nucleons and antinucleons increases, while that between kaons and
antikaons decreases, consistent with the results in
Ref.~\cite{Son12}. It is thus expected that further reducing the
value of $R_V$ would underestimate the relative $v_2$ difference
between nucleons and antinucleons and overestimate that between
kaons and antikaons, while further increasing the value of $R_V$
would underestimate that between kaons and antikaons. To reproduce
both the relative $v_2$ differences for nucleons and antinucleons as
well as that for kaons and antikaons requires the value of $R_V$ to
be within 0.5 and 1.1. According to
Refs.~\cite{Asa89,Fuk08,Car10,Wei12}, such values of vector coupling
would make the critical point disappear in the QCD phase diagram,
and the hadron-quark phase transition would always be a smooth
crossover. Furthermore, a larger value of $R_V$ results in a
hadron-quark phase transition at very high densities or even a
disappearance of the quark phase in neutron stars~\cite{Sha13},
leading to a possible explanation for the observed two-solar-mass
neutron star~\cite{Dem10}.

In summary, we have studied the effects of both the partonic and the
hadronic potential on the elliptic flow splitting of particles and
their antiparticles in relativistic heavy-ion collisions carried out
in the BES program at RHIC. With the evolution of the partonic phase
described by an NJL transport model, we have obtained a larger $v_2$
for nucleons and $K^-$ than antinucleons and $K^+$, respectively,
right after hadronization. After the hadronic evolution described by
a relativistic transport model that includes the empirically
determined hadronic potentials for particles and antiparticles, the
final $v_2$ is larger for nucleons and $K^+$ than antinucleons and
$K^-$, respectively. The relative $v_2$ differences from the STAR
data can be reproduced if the ratio $R_V$ of the vector coupling
constant $G_V$ to the scalar coupling constant $G$ in the NJL model
is between 0.5 and 1.1, after taking into account the mean-field
potential effects in both the partonic and the hadronic phase. This
result is expected to remain unchanged after including the effect of
gluons, e.g., using the PNJL model with the Polyakov loop
contributions, since it is similar for quarks and antiquarks and can
be compensated by using a different parton scattering cross section.
Our results therefore suggest that studying the $v_2$ splitting of
particles and their antiparticles in heavy-ion collisions provides
the possibility of studying the QCD phase structure at finite baryon
chemical potential, thus helping understand the nature of the strong
interaction.

\begin{acknowledgments}
One of us (Jun Xu) would like to thank Zi-Wei Lin for helpful
communications. This work was supported by the Major State
Basic Research Development Program in China (No. 2014CB845401), the
"Shanghai Pujiang Program" under Grant No. 13PJ1410600, the
"100-talent plan" of Shanghai Institute of Applied Physics under
grant Y290061011 from the Chinese Academy of Sciences, the US
National Science Foundation under Grant No. PHY-1068572, and the
Welch Foundation under Grant No. A-1358.
\end{acknowledgments}

\end{document}